\documentclass[aps,prb,twocolumn,superscriptaddress,longbibliography]{revtex4-1}

\usepackage[colorlinks=true,citecolor=blue]{hyperref}
\usepackage{graphicx}
\usepackage{amsmath}
\usepackage{amssymb}

\DeclareGraphicsExtensions{.png,.jpg,.eps}
\usepackage{xcolor}

\newcommand{\maf} {m_{\text{AF}}(\mathbf r)}
\newcommand{\swave} {\Delta(\mathbf r)}
\newcommand{\mafo} {m_{\text{AF}}}
\newcommand{\swaveo} {\Delta}
\newcommand{\valley} {\mathcal{V}}
\renewcommand{\H} {\mathcal{H}}

\begin{document}

\author{Senna S. Luntama}
\affiliation{Department of Applied Physics, Aalto University, 00076 Aalto, Espoo, Finland}

\author{P\"{a}ivi T\"{o}rm\"{a}}
\affiliation{Department of Applied Physics, Aalto University, 00076 Aalto, Espoo, Finland}

\author{Jose L. Lado}
\email{jose.lado@aalto.fi}
\affiliation{Department of Applied Physics, Aalto University, 00076 Aalto, Espoo, Finland}

\title{
	Interaction-induced topological superconductivity
	in antiferromagnet-superconductor junctions
}

\begin{abstract}
 
	We predict that junctions between
	an antiferromagnetic insulator and a superconductor provide
	a robust platform to create a one-dimensional
	topological superconducting
	state. Its emergence does not require the
	presence of intrinsic
	spin-orbit coupling nor non-collinear magnetism,
	but arises solely from repulsive electronic
	interactions on interfacial solitonic states.
	We demonstrate that a topological
	superconducting state is generated by repulsive interactions
	at arbitrarily small coupling strength, and
	that the size of the topological gap
	rapidly saturates to the one of the
	parent trivial superconductor.
	Our results put forward
	antiferromagnetic insulators
	as a new platform for interaction-driven
	topological
	superconductivity. 
\end{abstract}

\date{\today}

\maketitle


The search for topological superconductors has been one of the most
active areas in condensed matter
physics in recent years~\cite{Kitaev2001,Alicea2012,Beenakker2013,Beenakker2016,Deng2012,PhysRevLett.110.126406,PhysRevLett.100.096407,PhysRevLett.114.236803,PhysRevLett.107.097001,PhysRevLett.105.077001,PhysRevLett.112.217001,PhysRevLett.117.047001,PhysRevB.87.104513,He2017,PhysRevLett.120.116801,PhysRevLett.115.197204,NadjPerge2014,Albrecht2016,Deng2016}. These systems, pursued in
particular for the emergence of Majorana zero modes,
represent one of the potential
solid state platform for the
implementation of topological quantum computing~\cite{Alicea2011,PhysRevX.4.011036}.
Due to their elusive
nature, topological superconductors are often artificially engineered.
A variety of platforms have been proposed and demonstrated for
this purpose~\cite{NadjPerge2014,Mourik2012,2020arXiv200202141K}, 
generically relying on a combination of
conventional s-wave superconductivity, 
ferromagnetism and strong spin-orbit
coupling~\cite{PhysRevLett.100.096407,PhysRevLett.114.236803,PhysRevLett.107.097001,PhysRevLett.105.077001,PhysRevLett.120.116801,PhysRevLett.105.177002,PhysRevLett.104.040502}.

While ferromagnets have played a central role for 
artificial topological
superconductivity, antiferromagnetic insulators 
have been overlooked for this purpose.
Recently, antiferromagnets have attracted a great amount of attention
due to their unique properties for spintronics~\cite{Jungwirth2016,RevModPhys.90.015005,PhysRevLett.122.077203,PhysRevLett.113.196602,PhysRevB.102.140504} and
for creating novel types of topological matter~\cite{Otrokov2019,Liu2020,PhysRevLett.124.066401,PhysRevB.81.245209,PhysRevB.90.060507,PhysRevLett.124.197201,PhysRevLett.124.066401,PhysRevLett.125.037201}.
Ferromagnetism efficiently lifts Kramer's degeneracy,
a process heavily detrimental for spin-singlet superconductivity.
Antiferromagnetism, in comparison, does not lift
Kramer's degeneracy between opposite spins in the absence
of spin-orbit coupling, a feature that could potentially
make antiferromagnetism
more compatible with spin-singlet superconductivity.\cite{PhysRevB.22.2307,PhysRevB.80.064508,PhysRevLett.103.207002,PhysRevB.68.144517,PhysRevLett.46.614,PhysRevLett.121.157003,Coh2015,Wang2016}. 

\begin{figure}[!t]
    \centering
    \includegraphics[width=\columnwidth]{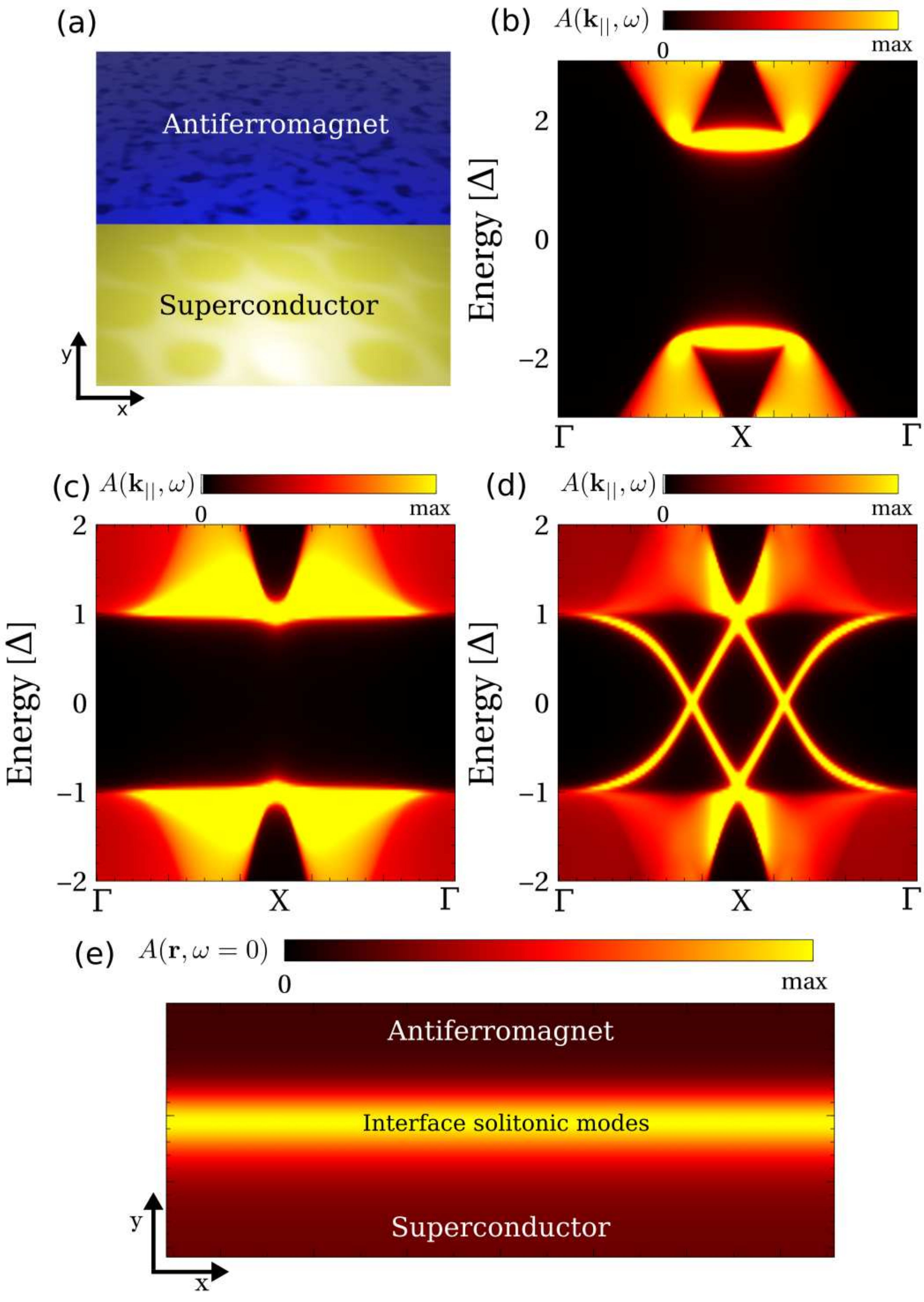}
	\caption{(a) A sketch of the two dimensional antiferromagnet (AF) and superconductor (SC) forming a one-dimensional AF-SC interface.
	The spectral function at the
	surface of the AF (b), at the surface of the SC (c)
	and at the interface between AF and SC (d) as given by our model Hamiltonian (\ref{eq:fullh}) in a honeycomb lattice.
	Panel (e) shows the spatial distribution of the interfacial modes.
	Here we chose $\swaveo=0.3t$, $\mafo=0.5t$,  $\mu=t$ and $V_1=V_2=0$.}
    \label{fig:fig1}
\end{figure}

Here we show that two-dimensional
topologically trivial
antiferromagnetic insulators
provide a platform to design
one-dimensional topological superconductivity.
In our proposal, spin-orbit coupling effects are not
necessary for topological superconductivity to appear,
nor a fine-tuning between the different components of the
system. In contrast,
we show that long-range interactions alone give rise to a non-trivially
gapped state hosting Majorana excitations, and
that the interaction-induced gap opening 
is topological irrespective of details. We demonstrate
that the robustness of this unique 
state stems from the
solitonic nature of the emergent excitations at the interface,
in which interaction-induced gap opening unavoidably gives rise
to a topological superconducting state.
Our results put forward
antiferromagnet-superconductor
junctions as a robust platform to engineer
interaction-induced topological superconductivity.

Our system consists of a junction between a conventional
s-wave superconductor and antiferromagnetic insulator,
as shown in Fig.~\ref{fig:fig1}a.
To model this system, we take a Hamiltonian
in the honeycomb lattice
of the form

\begin{equation}
	\H = \H_{\text{kin}} + \H_{\text{AF}} + \H_{\text{SC}} + \H_{\text{int}}
	\label{eq:fullh}
\end{equation}

where 
\begin{equation}
	\H_{\text{kin}}=
	t\sum_{\langle ij \rangle,s} 
	c^\dagger_{i,s} c_{j,s} +
	\sum_{i,s} \mu(\mathbf r_i) 
	c^\dagger_{i,s} c_{i,s} 
	\label{eq:mu}
\end{equation}

\begin{equation}
	\H_{\text{AF}}=
	\sum_{i,s} m_{\text{AF}}(\mathbf r_i) 
	\tau^z_{i,i} \sigma^z_{s,s}
	c^\dagger_{i,s} c_{i,s} 
	\label{eq:af}
\end{equation}

\begin{equation}
	\H_{\text{SC}}=
	\sum_{i} \Delta (\mathbf r_i) 
	c_{i,\uparrow} c_{i,\downarrow}  + \text{H.c.}
	\label{eq:sc}
\end{equation}

\begin{equation}
	\begin{split}
		\H_{\text{int}} & =
	V_1
	\sum_{\langle i j \rangle } 
	\left ( \sum_s 
	c^\dagger_{i,s} c_{i,s}
	\right )
	\left ( \sum_s 
	c^\dagger_{j,s} c_{j,s}
	\right ) 
	+ \\
		& V_2
	\sum_{\langle \langle i j \rangle \rangle }
	\left ( \sum_s 
	c^\dagger_{i,s} c_{i,s}
	\right )
	\left ( \sum_s 
	c^\dagger_{j,s} c_{j,s}
	\right ) 
	\end{split}
	\label{eq:hint}
\end{equation}
where 
$c^\dagger_{i,s}$
is the fermionic creation operator
for site $i$ and for spin $s$,
$\sigma^z$ denotes the spin Pauli matrix,
$\tau^z$ the sublattice Pauli matrix,
$\langle \rangle$ the first neighbors and
$\langle\langle \rangle\rangle$ the second neighbors.
Taking that the interface between the antiferromagnet and the
superconductor is located at $\mathbf r = (x,0,0)$
we take 
$\Delta (\mathbf r) = \frac{\Delta}{2}[1-\text{sign}(y)]$
$\mu (\mathbf r) = \frac{\mu}{2}[1-\text{sign}(y)]$
and
$m_{\text{AF}} (\mathbf r) =
\frac{\mafo}{2}[1+\text{sign}(y)]$\footnote{Taking a smooth
transition between the superconductor and
antiferromagnet does not impact the results qualitatively}.
The repulsive
interaction term of Eq.~\ref{eq:hint}
is solved at the mean-field level including the usual
mean-field decouplings
$\H_{\text{int}} \approx \H^{\text{MF}}
= \sum \chi_{ijss'} c^\dagger_{i,s} c_{j,s'}$
with $\chi_{ijss'}$ the self-consistent mean-field parameters
\footnote{Charge renormalization is reabsorbed in Eq.~\ref{eq:mu}}.
On-site interactions are incorporated in $\maf$ and $\swave$ at the mean-field level.

It is instructive to examine the electronic bandstructure
in the absence of interactions and in the absence of an interface.
Let us consider a semi-infinite slab in the
$y-$direction, having translational symmetry in the
$x-$direction as depicted in Fig.~\ref{fig:fig1}a.
For that geometry, we compute the
momentum-resolved spectral function at the
edge $A(\mathbf{k}_{||},\omega) = - \frac{1}{\pi}
\text{Im} \left (\omega - \H(\mathbf{k}_{||})+i0^+\right )^{-1}$ using the Dyson formalism~\cite{Sancho1985}.
For both isolated 
superconductor and antiferromagnet, the
surface spectral function presents a gap, as shown in
in Fig.~\ref{fig:fig1}bc, that simply stems from
the gapped topologically trivial band structure. 
In the case of the superconductor the
gap is controlled by $\swaveo$, whereas in the
antiferromagnet, the gap is determined by
$\mafo$. In stark contrast, when the antiferromagnet and superconductor
are joined together, a new branch of interfacial modes
appear as shown in Fig.~\ref{fig:fig1}d.
By computing the spectral function in real space
at zero energy $A(\mathbf r,\omega=0)$ it is
clearly seen that the new branch is
heavily localized
at the junction between the
superconductor and the antiferromagnet. 
We have verified, that for different values of the
superconducting and antiferromagnet order 
parameters, zero modes emerge as long as the order parameters are not
substantially bigger than the typical bandwidth. 

The emergence of the 
interfacial zero modes can be rationalized
from a low energy model for the honeycomb lattice~\cite{PhysRevD.13.3398,PhysRevX.5.041042,PhysRevLett.121.037002,PhysRevB.100.125411,PhysRevResearch.2.023347}.
For the following analytic derivation, it is convenient to take
$\mu=0$ so that the full antiferromagnet-superconductor
can be described
with a generalized Dirac equation
at the $K$ point of the honeycomb lattice~\cite{RevModPhys.81.109}.
The low energy excitations can be captured
by an effective model around the valleys $\valley^z=\pm 1$,
and we will focus first on
taking the momentum parallel to the
interface $p_x=0$.
By defining
the Nambu spinor $\Psi^\dagger = (
c^\dagger_{A,\uparrow,\mathbf k},
c^\dagger_{B,\uparrow,\mathbf k},
c_{A,\downarrow,-\mathbf k},
c_{B,\downarrow,-\mathbf k}
)$, the Hamiltonian
in the electron-up/hole-down
sector ($\Uparrow$)
can be written
as $\mathcal{H} (p_x=0,p_y)_\kappa= \frac{1}{2}\Psi^\dagger H_\kappa \Psi$ with

\begin{figure}[t!]
    \centering
    \includegraphics[width=\columnwidth]{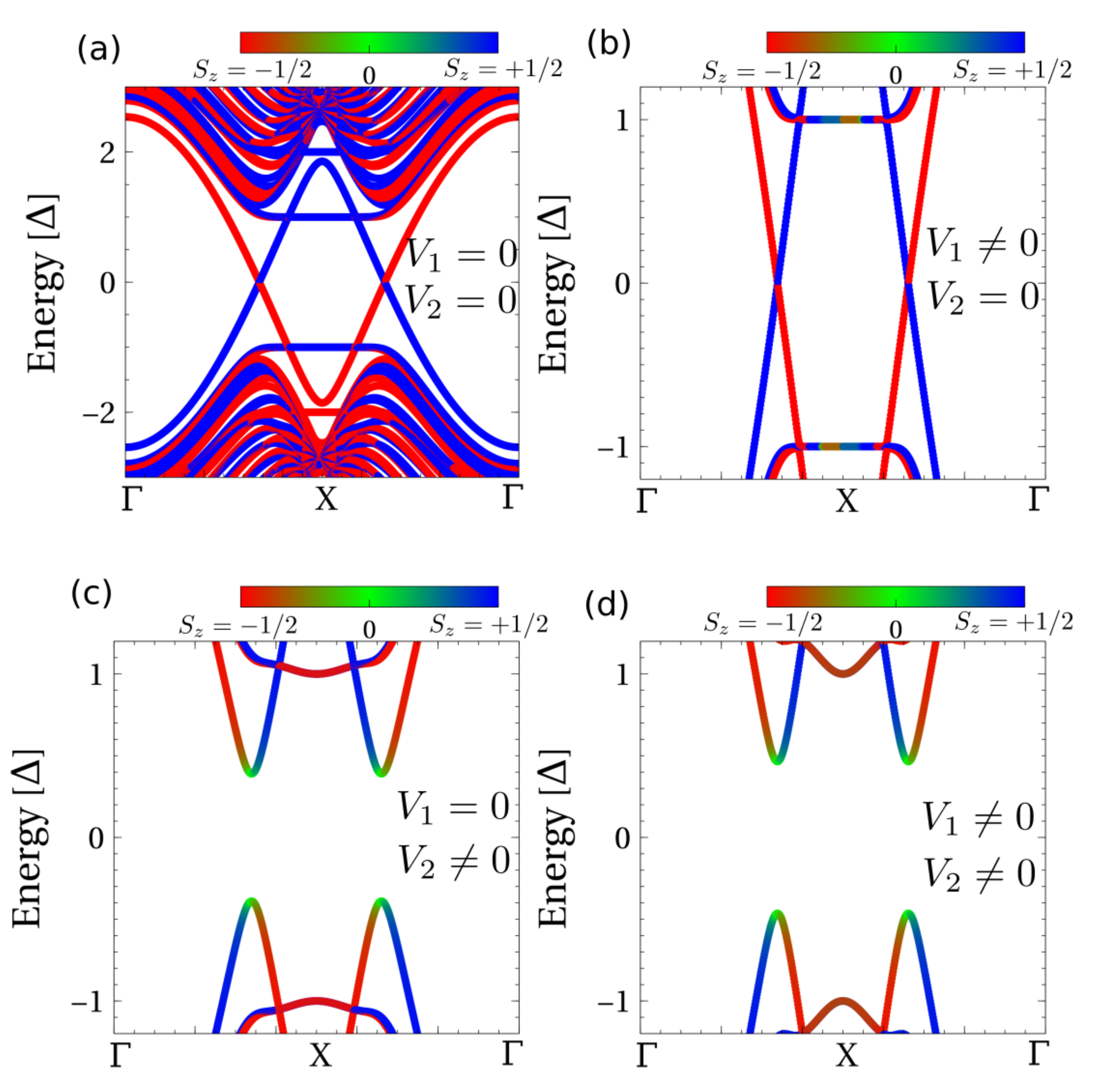}
        \caption{(a) Non-interacting bands in a ribbon
	geometry. First
	neighbor interactions do not lead to a gap (b),
	whereas second neighbor interactions
	drive a gap opening (c).
	When both first and second neighbor
	interactions are present
	the gap remains.
	The parameters are $V_1=t$ in (b), $V_2=1.7t$ in (c)
	$V_1=t$ $V_2=2t$ in (d) and $m_{AF}=0.8t$
	$\Delta=0.4t$ in (a-d). 
	}
    \label{fig:fig2}
\end{figure}

\begin{equation}
	H_\kappa = 
	\begin{pmatrix}
		\maf & p_y & \swave & 0 \\
		p_y & -\maf & 0 & \swave \\
		\swave & 0  & \maf & -p_y & \\
		0 & \swave & -p_y & -\maf .\\
	\end{pmatrix}
\end{equation}
The spectrum of this
effective model is gapped
at $y\pm \infty $, as expected
from its asymptotic
antiferromagnet/superconductor gap.
However, a zero energy mode
$H | \psi_\Uparrow \rangle = 0$
at the interface can be always
built, taking the functional form
$\psi_\Uparrow^\dagger = 
e^{-\int_0^y [\Delta (y') - m_{\text{AF}} (y') ] d y'}
(
c^\dagger_{A,\uparrow}+i
c^\dagger_{B,\uparrow}-i
c_{A,\downarrow}-
c_{B,\downarrow}
)$. 
The nature of this zero mode is analogous to the 
Jackiw-Rebbi soliton~\cite{PhysRevD.13.3398},
and therefore can be understood as an antiferromagnet-superconducting
soliton.
The complementary
electron-down/hole-up ($\Downarrow$) sector
of the Hamiltonian will therefore also host a
zero mode, that we label as $\psi_\Downarrow$. 
Away from the point $p_x=0$, the previous
state acquires a finite dispersion given by first
order perturbation theory 
$v_F p_x = \langle \psi_\Uparrow | \mathcal{H} | \psi_\Uparrow \rangle$. 
As a result, close to the
$K$-points two branches of zero modes
appear, giving rise to the effective
low energy Hamiltonian
\begin{equation}
	H (p_x)= \sum_\kappa v_F p_x \valley^{z}_{\kappa,\kappa} [
		\psi^\dagger_{\Uparrow,\kappa,p_x} \psi_{\Uparrow,\kappa,p_x}
	-
		\psi^\dagger_{\Downarrow,\kappa,p_x} \psi_{\Downarrow,\kappa,p_x}
		]
		\label{eq:heff}
\end{equation}
where $\kappa$ runs over the two valleys.
It is interesting to note that the four modes are not independent,
but they are related by electron-hole symmetry operator
$\Xi = \theta^y \sigma^y \mathcal{C}$ with $\theta^y$ the Nambu Pauli matrix and
$\mathcal{C}$ complex conjugation 
as
$\Xi^{-1} \psi_{\Uparrow,+1,p_x} \Xi = \psi_{\Downarrow,-1,-p_x}$
due to the built-in Nambu
electron-hole symmetry of the Hamiltonian.
Therefore, the Hamiltonian Eq.~\ref{eq:heff}
hosts only two physical degrees of freedom, each one propagating in opposite
directions, realizing an effective spinless one-dimensional model.
These singly-degenerate
channels are analogous
to quantum Hall edge states~\cite{PhysRevLett.120.116801},
and helical channels in topological
insulators~\cite{PhysRevLett.100.096407},
states that provide a starting point
for engineering a
topological superconducting gap. 
Remarkably in our case, as will be shown below, the solitonic 
gapless channels
will open up a topological superconducting
gap once electron-electron
interaction effects are included.

Let us now move on to consider the impact of long-range
electronic interactions in the solitonic modes. 
For computational convenience, we now perform our calculations in ribbons
of finite width in the $x$-direction, in which we take the transverse direction
wide enough to avoid finite-size effects.
The previous gapless interface modes of Fig.~\ref{fig:fig1}d
and derived in Eq.~\ref{eq:heff} appear in this ribbon
geometry as shown in Fig.~\ref{fig:fig2}a,
where
$S_z = \frac{1}{2}\langle \sum_{n,s} \sigma^z_{s,s} c^\dagger_{n,s} c_{n,s} \rangle_{\Psi_k} $ with $\Psi_k$ the eigenstate.
It is shown that in the absence of interactions, the sectors $S_z = \pm 1/2$
are fully decoupled, stemming from the $U(1)$-spin symmetry
of the Hamiltonian.
With this lattice model, we now explore the impact
of electronic interactions by solving
self-consistently Eq.~\ref{eq:fullh}. Note that the interactions apply both along the interface and across it. 
We start by considering only first
neighbor
interactions, taking $V_2=0$. In this situation, a gap does not open even when $V_1$ is increased,
as shown in Fig.~\ref{fig:fig2}b. 
We now move on to the case of $V_2$, taking first
$V_1=0$. As observed in Fig.~\ref{fig:fig2}c, it is clearly seen that now a gap opens up.
This behavior also takes place when $V_1$ is taken to be non-zero, see Fig.~\ref{fig:fig2}d. As a result, second neighbor interactions are the
only interaction capable of opening up a gap on the topological
interface modes,
whose magnitude is marginally affected by the first neighbor interactions.

\begin{figure}[t!]
    \centering
    \includegraphics[width=\columnwidth]{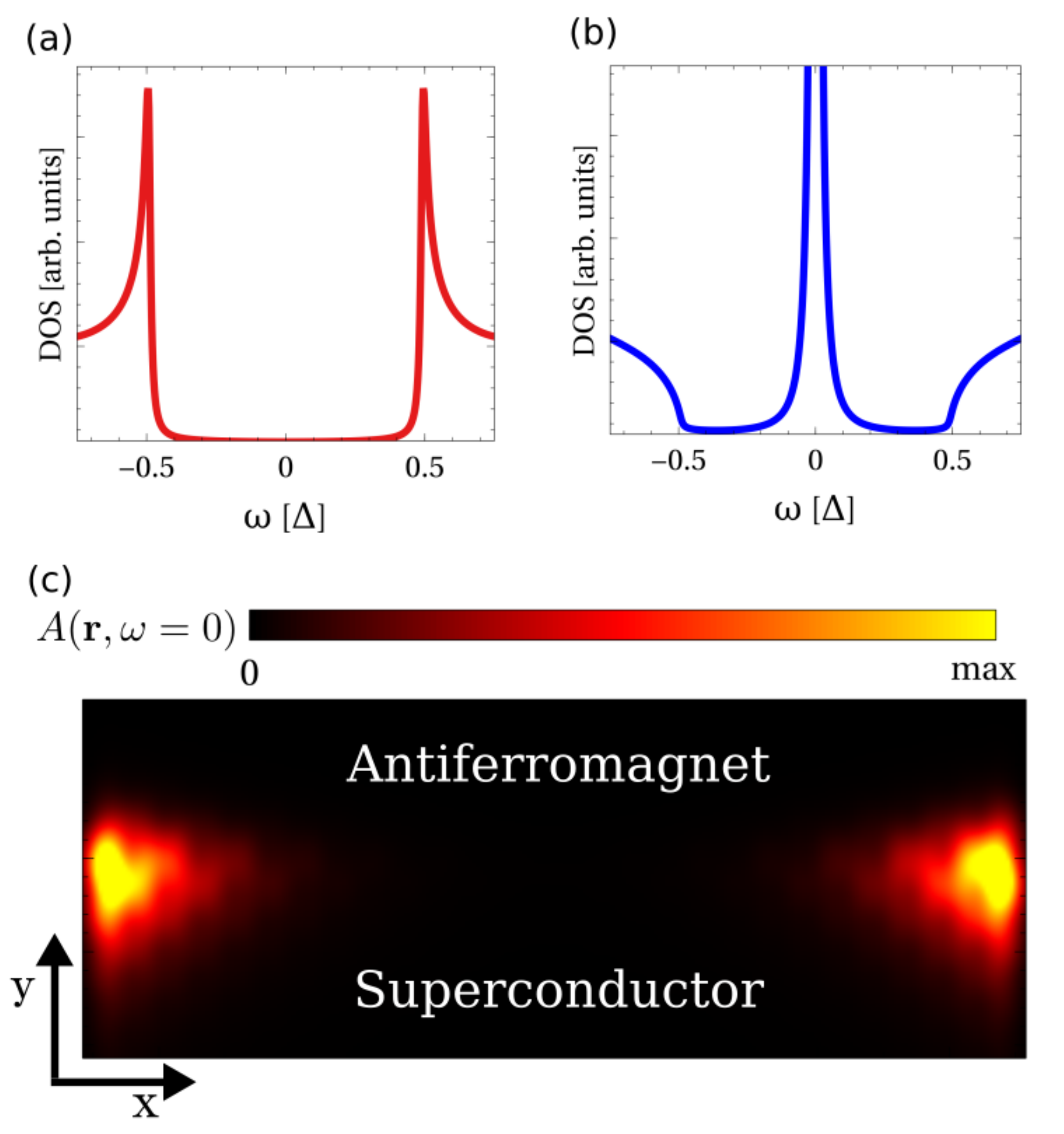}
        \caption{(a) Spectral function
	in the bulk in the presence of interactions, and
	(b) at the edge showing the emergence of a zero
	Majorana mode.
	Panel (c) shown the spectral function at $\omega=0$
	for a finite junction, featuring edge zero
	Majorana modes.
	We used now $\swaveo=0.4t$, $\mafo =0.8t$,
        $V_1=t$ and $V_2 =2t$.
        }
    \label{fig:fig3}
\end{figure}

The emergence of a gap opening driven by electronic interactions
raises the question of potential non-trivial topological properties. 
From the point of view of the effective
low energy model, interactions create
an effective term in Eq.~\ref{eq:heff}
of the form
$H^{MF} \sim 
\langle \Psi_\Uparrow \Psi^\dagger_\Downarrow \rangle 
\Psi^\dagger_\Uparrow \Psi_\Downarrow + \text{H.c.}
$. 
It is interesting to note that due to the solitonic functional
form of $\Psi_\Uparrow$ and $\Psi_\Downarrow$ and their relation
via electron-hole symmetry, the gap ($\propto \langle \Psi_\Uparrow \Psi^\dagger_\Downarrow \rangle$) created is odd with respect
to $\kappa$, the valley index, suggesting the emergence of an effective topological
superconducting state.
To verify the non-trivial topological nature of the interaction-driven
gapped state, we compute both its $Z_2$ topological invariant~\cite{Kitaev2001,PhysRevB.88.075419} and surface spectral
function. 
We revealed that the gapped system has a topologically non-trivial
$Z_2$ invariant, signaling the existence of a topological superconducting state.
This is further verified when computing the density of states at the edge
of the interface in a ribbon that spans from $x=0$ to $x=\infty$,
as shown in Fig.~\ref{fig:fig3}a. 
The edge of the system hosts a zero-mode resonance associated with the
unpaired Majorana stemming from the non-trivial electronic structure.
This is contrasted with the finite gap present in the bulk of the system shown in
Fig.~\ref{fig:fig3}b. The localization of the zero-mode can also be seen when computing
the spectral function for $\omega=0$, Fig.~\ref{fig:fig3}c.

Let us now move on to look at the impact of long-range interactions, and in particular,
at the interplay between the first and second neighbor interactions at the mean field level. 
For the sake of simplicity in the following discussion
we will only consider effects that appear by means of a mean field
decoupling of Eq.\ref{eq:hint}, without considering
beyond mean-field effects
or additional $t-J$ contributions.
At the
mean-field level, the
interaction term of Eq.~\ref{eq:hint} can give rise to two potential effects: first,
interaction induced hoppings and second,
symmetry broken states such as charge
density waves. In the weak
coupling regime considered here, only
interaction-induced hopping terms arise.
In particular, the
time-reversal symmetric
and spin-dependent
part of 
$\chi_{ijss'}$
yield an effective
spin and spatially-dependent
synthetic spin-orbit coupling
term of the Kane-Mele form~\cite{PhysRevLett.100.096407}
\footnote{Details on the microscopic
mean-field
are included in the Supplemental Material}.
This interaction-induced
term creates spin-mixing in the
solitonic modes, opening up a topological gap.

The interplay of first and second neighbor interactions can be easily rationalized
within this language. From the mean-field point of view, first neighbor interactions
can give rise to interaction induced Rashba spin-orbit coupling terms~\cite{PhysRevB.82.161414}, whereas
second neighbor interactions can give rise to interaction-induced Kane-Mele spin-orbit
coupling~\cite{PhysRevLett.95.226801}. However, due to valley polarized nature of the solitonic modes, interaction
induced Rashba-spin-orbit coupling does not open up a gap in them~\cite{PhysRevB.82.161414},
whereas Kane-Mele like spin-orbit~\cite{PhysRevLett.95.226801} 
can create a gap. As a result, second
neighbor interactions are the only ones capable of interaction-induced
gap opening in the system.
In contrast, the effect of the first neighbor interactions is to simply create a
Fermi velocity renormalization~\cite{PhysRevB.88.205429,PhysRevLett.118.266801} increasing the kinetic energy
of the solitonic modes, yet without any competing mechanism for gap opening.

\begin{figure}[t!]
    \centering
    \includegraphics[width=\columnwidth]{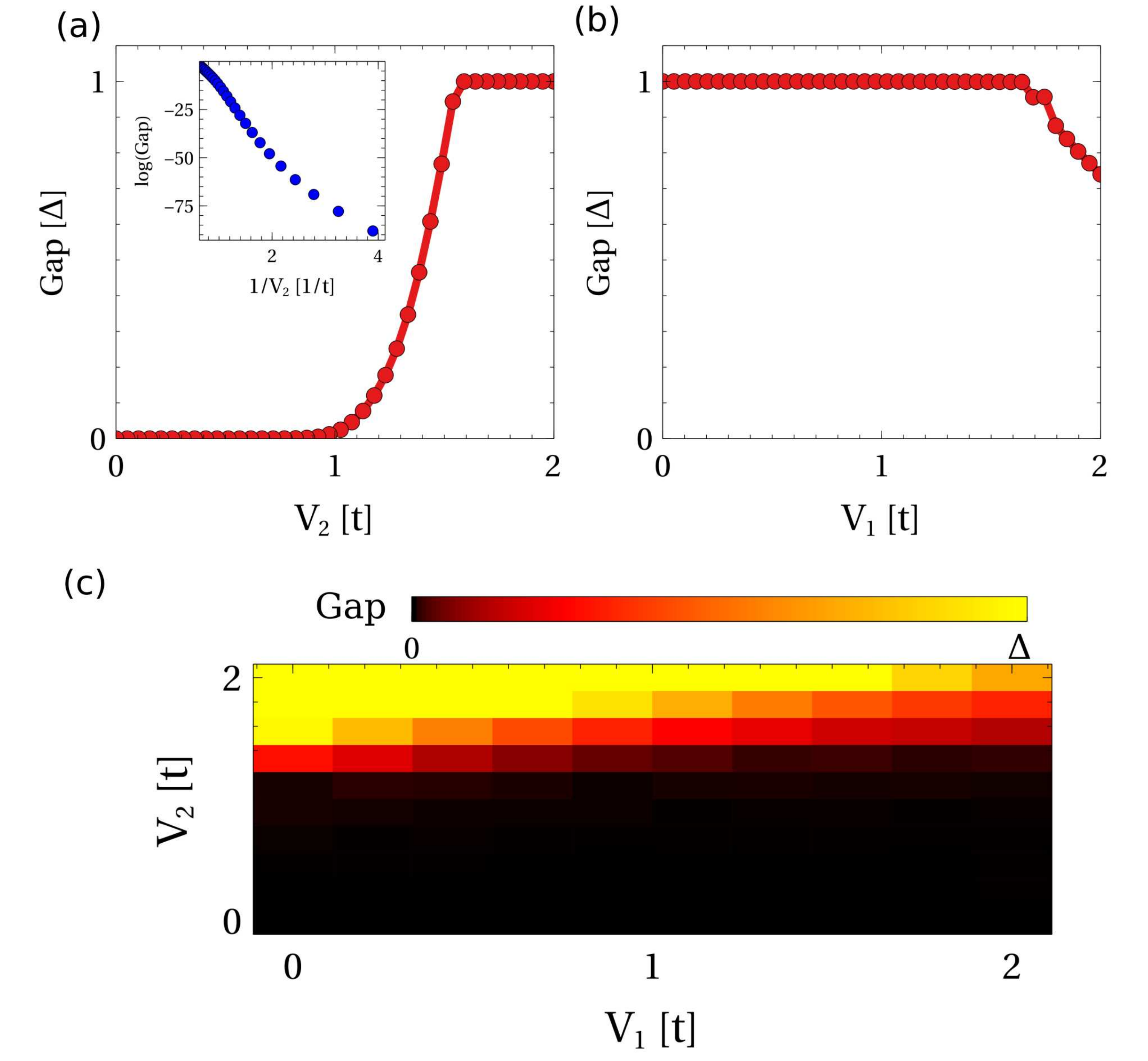}
        \caption{(a) 
	Evolution of the
	topological gap
	with the
	electron-electron interaction: (a)
	as a function of 
	$V_2$ taking
	$V_1=0$, (b) as a function of
	$V_1$ taking $V_2=2t$. 
	Panel (c) shows the
	topological gap as a function
	of the two electronic interactions 
	$V_1$ and $V_2$, highlighting that
	only the second neighbor interaction
	opens up a gap.
	We took $\Delta=0.2t$ and $\mafo =0.4 t$.
        }
    \label{fig:fig4}
\end{figure}

It is crucial to understand whether the gap opening requires a finite minimum value of interaction strength.
We investigate this by taking the first neighbor interaction $V_1=0$,
and looking at the topological gap as a function of the repulsive
second neighbor interaction
$V_2$. It is clearly observed that the topological gap becomes stronger
as $V_2$ is increased, without the existence of a critical value for the
transition (Fig.~\ref{fig:fig4}a). 
In particular, a logarithmic plot of the gap (inset of Fig.~\ref{fig:fig4}a)
at small coupling strength
reveals that the topological gap $\delta$ follows an
exponential dependence $\delta \sim e^{-\frac{v_F}{V_2}}$
\footnote{The term in $\H^{MF}$
creating the gap opening
shows a similar
exponential dependence.}.
Interestingly,
whereas exponential dependences of that form are typical for
superconducting instabilities driven by attractive interactions~\cite{PhysRev.108.1175},
in our present case interactions are actually repulsive.
This behavior stems from the projection of the interactions
in the low energy solitonic model of Eq.~\ref{eq:heff},
driving a topological phase transition
at arbitrarily small couplings.
At large coupling strengths $V_2$, the topological gap saturates to the gap
of the superconductor. This behavior should be contrasted
with the other schemes proposed for topological superconductivity, in which
the topological gap is usually substantially smaller than the original
superconductor gap.
This saturation of the topological gap can be ascribed to the
absence of competition between the superconductor and the antiferromagnet.
Including finite
first neighbor interactions $V_1$ keeps the picture qualitatively
unchanged, yet with a slightly renormalized topological
gap (Fig.~\ref{fig:fig4}b).
The interplay between $V_1$ and $V_2$ shown in Fig.~\ref{fig:fig4}c shows that
whereas $V_2$ opens the topological gap, $V_1$ leaves the system gapless
or slightly renormalizes the topological gap.
Finally, we note
that imperfections and disorder
are known to potentially
impact topological
superconductors by limiting the localization length
and reducing the topological gap~\cite{PhysRevLett.109.146403,PhysRevB.93.075129,PhysRevB.89.144506}.
We verified that the phenomenology presented above
is resilient towards Anderson
disorder 
and happens for generic AF-SC
interfaces\footnote{Details on the role of Anderson disorder and
different interfaces are provided in the Supplemental Material.}.
Disorder slightly decreases the topological
gap, yet without qualitatively impacting our results.

Finally, we address the potential experimental
realization of our proposal.
For a solid-state realization,
no specific requirements
are necessary for the superconductor
besides conventional s-wave pairing,
as realized in NbSe$_2$. The fundamental requirement is having a two-dimensional
honeycomb antiferromagnetic insulator
\footnote{Although
square lattice models can have antiferromagnetism, 
their low energy model is usually not of the Dirac form.},
as its electronic structure is expected to have
the gapped Dirac points required for the emergence
of the topological solitonic modes.
Within van der Waals materials,
trihalides host 
a magnetic honeycomb lattice~\cite{McGuire2017},
and in particular 
antiferromagnetic 
strained trihalides~\cite{PhysRevB.98.144411,Wu2019}
would be suitable
for our proposal.
This pathway would require
creating
superconductor/antiferromagnet
devices with those strained van der Waals materials.
Within oxides,
thin films of
InCu$_{2/3}$V$_{1/3}$
O$_3$~\cite{PhysRevB.78.024420}
or 
$\ensuremath{\beta}{\text{-Cu}}_{2}{\text{V}}_{2}{\text{O}}_{7}$~\cite{PhysRevB.82.144416}
has the required
antiferromagnetic honeycomb lattice.
For this possibility, a single layer 
of the bulk oxide should
be epitaxially grown.
Generic 
two-dimensional antiferromagnetic insulators
hosting Dirac points in their
normal state~\cite{Wehling2014}
would be suitable materials
for our proposal, whose specific
$V_1, V_2$ parameters can be inferred by first principles methods~\cite{PhysRevLett.119.056401,PhysRevLett.111.036601,PhysRevB.98.085127,2020arXiv201103271T}
\footnote{A discussion on the $V_2$ values in two-dimensional
materials is provided in the Supplemental Material.}.
Finally, future ultracold atom setups~\cite{torma_quantum_2015}
are potential platforms for the
realization of our model,
as honeycomb structures~\cite{Jotzu2014},
antiferromagnetic correlations~\cite{Mazurenko2017},
long-range interactions\cite{Ferlaino2016,2020arXiv201005871G,Yan2013}
and
s-wave correlations~\cite{Mitra2017} in the normal state have been separately demonstrated. Interactions can be tuned from attractive to repulsive by magnetic fields; spatially dependent fields could be one way of creating the AF-SC interface, once superfluid correlations in a lattice have been reached.

To summarize, we have shown that an interface between a topologically trivial two-dimensional
superconductor and  antiferromagnetic
insulator gives rise to a one-dimensional solitonic gas. 
Upon introduction of repulsive long-range interactions, we have demonstrated
that a topological gap gets generated, giving rise to Majorana
zero energy
modes. The emergence of topological superconductivity
appears in the absence of intrinsic spin-orbit coupling
and is driven by repulsive Coulomb interactions.
We showed that the topological gap appears
at arbitrarily small interactions, and rapidly saturates
to the gap of the parent superconductor, in stark contrast
with conventional proposals involving competition
between ferromagnetism and superconductivity.
Our results propose a new mechanism to generate topological
superconductivity based on interacting
solitons, putting forward antiferromagnetic insulators as
a potential materials platform for Majorana physics.

\textbf{\textit{Acknowledgments:}}
We acknowledge
the computational resources provided by
the Aalto Science-IT project.
J. L. L. acknowledges
financial support from the
Academy of Finland Projects No.
331342 and No. 336243.
P.T.
acknowledges support by the Academy
of Finland under project numbers 303351, 307419, and
327293.

\section*{Appendix}

\appendix

\renewcommand{\thefigure}{S\arabic{figure}}
\renewcommand{\thesection}{S\arabic{section}}
\renewcommand{\theequation}{S\arabic{equation}}

In this Appendix, we show that an interaction-induced
topological gap appears for generic interfaces
(Sec. \ref{sec:inter}), we study the
role of disorder on the topological gap
(Sec. \ref{sec:dis}), analyze the microscopic origin
of the interaction induced spin-orbit coupling
(Sec. \ref{sec:soc}), and comment
on real material estimates of the interaction
(Sec. \ref{sec:v2}).

\section{Topological
superconductivity in generic AF-SC interface}
\label{sec:inter}

\begin{figure}[!t]
    \centering
    \includegraphics[width=\columnwidth]{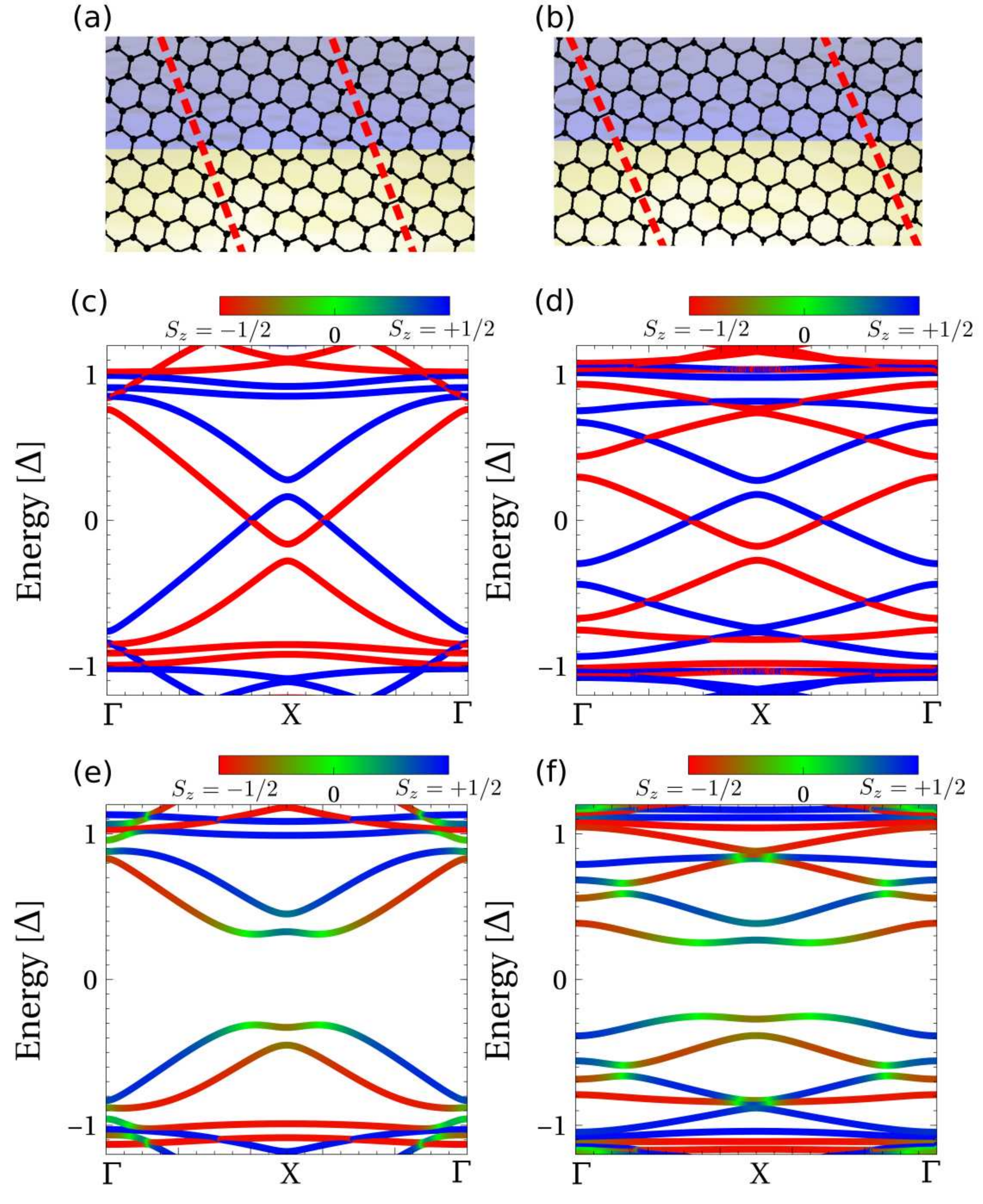}
        \caption{(a,b) Sketch of two different interfaces, where
        dashed lines denote the size of the unit cell.
        Panels (c,d) show the non-interacting band-structures,
        featuring the solitonic interface modes.
        Panels (e,f) show the band-structures once interactions are included,
        showing the emergence of a topological gap.
        We chose $\swaveo=0.2t$, $\mafo=0.4t$, and $V_2=1.7t$.}
    \label{fig:SMfig2}
\end{figure}

In the calculations included in the main manuscript, the orientation
of the interface is the zigzag one.
Here we show that generic
orientations of the interface would work for our proposal.
First, it is worth to mention that
the emergence of the solitonic modes is
not protected by a specific
lattice symmetry, but they arise at the interface between
inequivalently gapped Dirac equations. Those modes only appear
when the antiferromagnet is in contact with the superconductor,
and they are absent otherwise. The linear dispersion of the
modes is then obtained by perturbation theory to the
solitonic states. This phenomenology is expected to
appear in generic interfaces, suggesting the
emergent topological superconductivity does not depend on the
details of the interface.
In particular, we show in Fig. \ref{fig:SMfig2} the results
for a heterostructure with two different non-zigzag
interfaces. The specific structures are shown in
Fig. \ref{fig:SMfig2}ab, the non-interacting band-structures in
Fig. \ref{fig:SMfig2}cd, and the interacting mean-field
band-structures in Fig. \ref{fig:SMfig2}ef.
In particular, it is observed that in the absence of interactions,
the solitonic modes appear in generic interfaces. Furthermore,
when interactions are included, a topological gap opens up for both
interfaces. This phenomenology highlights the robustness
of the solitonic modes to the details of the
interface, and the generic emergence of a topological
state driven by interactions.

\begin{figure}[!t]
    \centering
    \includegraphics[width=\columnwidth]{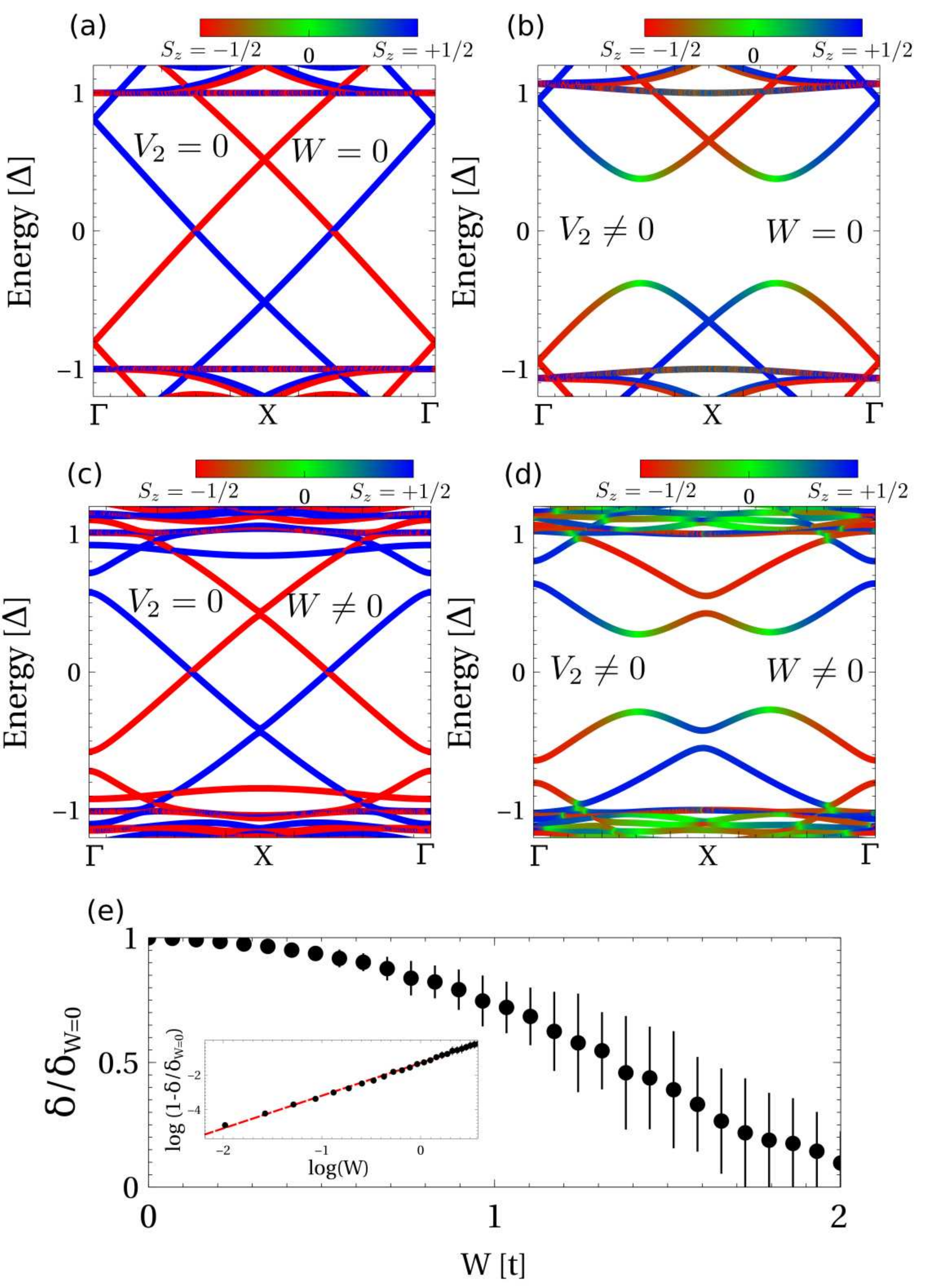}
        \caption{(a,b) Band-structure for a supercell
	of size $5$ in the absence of disorder (a,b) and
	in the presence of Anderson disorder $W=t$ (c,d).
	Panels (a,c) show the band-structure in the absence
	of interactions, and panels (c,d) in the presence
	of interactions. It is observed
	that the presence of disorder does not destroy
	the topological gap. Panel (e) shows the
	evolution of the ratio
	of the topological gap $\frac{\delta(W)}{\delta(W=0)}$
	as a function of the
	disorder strength $W$,
	averaged over 100 disorder configurations
	for each $W$.
	The inset of (e) shows a log-log plot,
	highlighting the power-law behavior.
	We chose $\swaveo=0.2t$, $\mafo=0.4t$, and $V_2=1.7t$ in (b,d,e).}
    \label{fig:SMfig3}
\end{figure}

\section{Impact of disorder}
\label{sec:dis}

The robustness to disorder
is one of the crucial points of any proposal
for Majorana states\cite{PhysRevLett.109.146403,PhysRevB.93.075129,PhysRevB.89.144506}. 
Since the fundamental physics of our proposal comes from the
interface modes, 
for the sake of concreteness
here we will in the following focus on
disorder effects that affect the interface.

\begin{figure}[t!]
    \centering
    \includegraphics[width=\columnwidth]{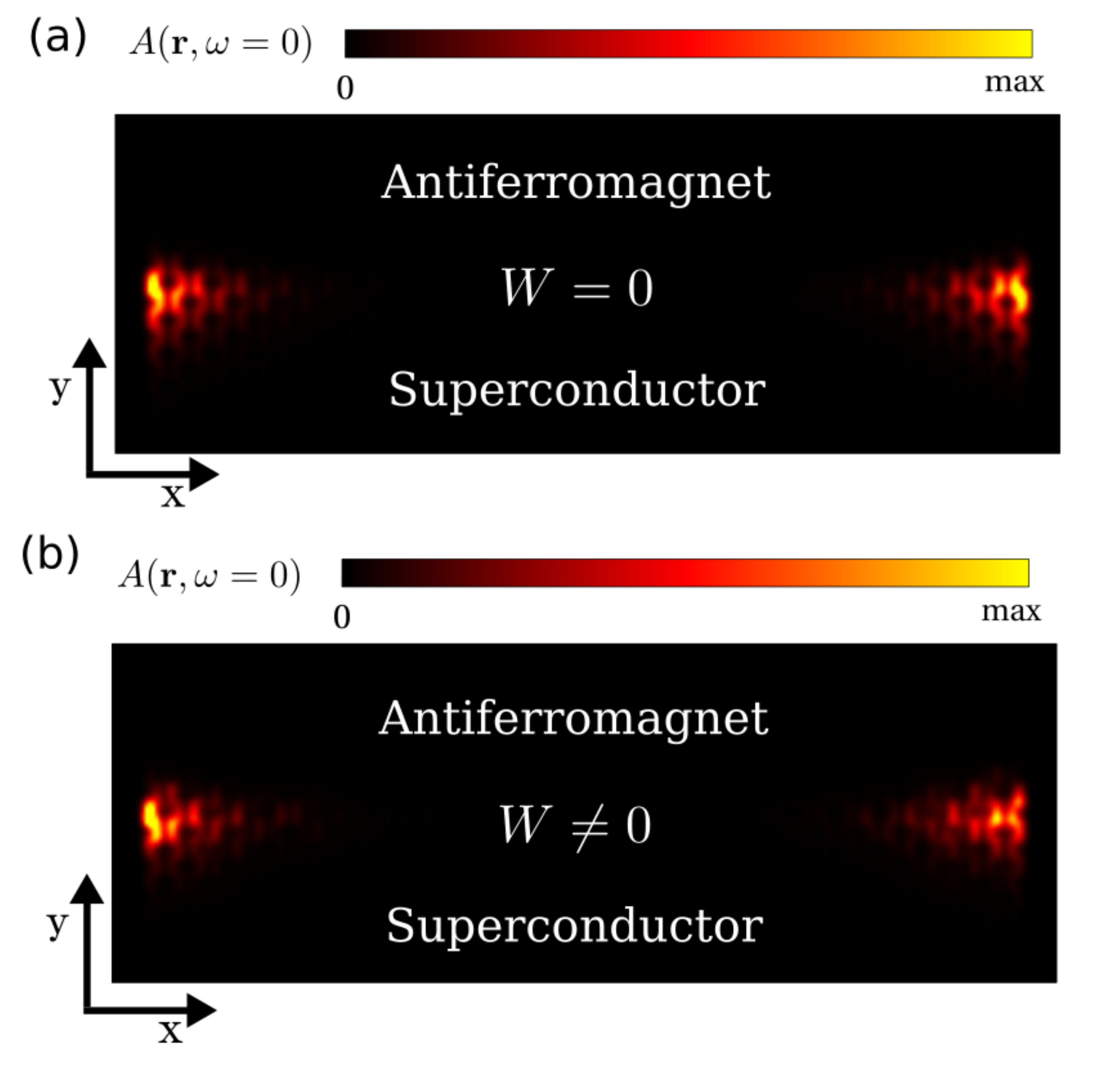}
        \caption{Local density of states
	at zero energy for a finite slab without
	disorder (a) and with disorder $W=t$ (b).
        We observe that Majorana zero modes are present in
	both cases, highlighting the robustness of the
	topological state to Anderson disorder.
        We chose $\swaveo=0.2t$, $\mafo=0.4t$, and $V_2=1.7t$.}
    \label{fig:SMfig4}
\end{figure}

First, we address the impact of Anderson disorder in a periodic
supercell in the $x-$direction.
For this goal we take a supercell of $5$ in the $x-$direction,
include Anderson disorder by adding a term to the Hamiltonian
of the form
\begin{equation}
	\H_W = \sum_{i,s} W_i c^\dagger_{i,s} c_{i,s}
\end{equation}
where $W_i$ is a random number between $[-W,W]$. The pristine case
corresponds to taking $W=0$. We show in Fig. \ref{fig:SMfig3}
the comparison between the electronic structures with and
without Anderson disorder in such supercell.
In particular, for the pristine supercells
we observe solitonic gapless modes in the absence of interactions
(Fig. \ref{fig:SMfig3}a), and a topological gap
in the presence of interactions (Fig. \ref{fig:SMfig3}b).
To demonstrate the robustness of our phenomenology,
we will consider a case with a relatively strong
disorder $W=t$.
When Anderson disorder is turned on, 
we observe than the solitonic modes remain mostly
unaffected (Fig. \ref{fig:SMfig3}c), and that in the
presence of interactions, a topological gap remains
(Fig. \ref{fig:SMfig3}d).
Interestingly, even with this strong disorder $W=t$, the
topological gap keeps 73\% percent of its pristine magnitude.
The reduction of the topological gap as a function of the disorder
is systematically explored in Fig. \ref{fig:SMfig3}e.
In particular, we observe that for a modest amount of disorder
$W=0.3t$, the topological gap keeps 97\% of its pristine value.
We have also performed a log-log plot
of the gap reduction (inset of Fig. \ref{fig:SMfig3}e),
getting that the disorder dependence of the gap follows
$\delta(W)/\delta(0) = 1 - (W/W_C)^\gamma$
with $\gamma \approx 2$, where $W_C$
is the critical disorder for the transition.
The robustness of the topological state
can be rationalized from its
impact on the parent electronic structures. First, the s-wave
superconducting state is resilient towards Anderson disorder
as follows from Anderson's theorem\cite{Anderson1959,Andersen2020}. The antiferromagnetic insulator
is also robust to disorder due to its antiferromagnetic gap.
And ultimately, the interface states are resilient to Anderson disorder
due to their solitonic nature\cite{PhysRevD.13.3398}. Finally, since the topological
gap stems from an interaction-induced gap opening of the
interface modes, the impact of the disorder on the topological
gap is small due to the robustness of the solitonic modes.

Finally, we consider the effect of Anderson disorder in a large
finite system, and in particular
its impact on the Majorana zero-edge modes. For this sake, we now
create a large system formed of 60 bulk cells in the $x-$direction,
and we compute the density of states at zero energy.
This is compared for the case without disorder $W=0$, and for the
case with Anderson disorder $W=t$, as shown in Fig. \ref{fig:SMfig4}.
We observe that in both cases, Majorana zero modes are located at the
left and right ends of the interface. In particular, the persistence
of zero modes in the disordered case highlights the robustness of the
topological state towards Anderson disorder. Its origin can be rationalized
as in the bulk case presented above.

\section{Interaction induced spin-orbit coupling}
\label{sec:soc}

Here we comment on the specific form of the interaction-induced
synthetic spin-orbit coupling.
First, it is worth emphasizing that generically, onsite interactions
$U$ will also appear in the honeycomb model. These interactions
are effectively included in our model at the mean-field level by means
of the antiferromagnetic field.
The existence
of this antiferromagnetic order quenches any potential charge
density wave orders or Haldane/Kane-Mele phases induced by $V_1$ and $V_2$.
Therefore, both $V_1$ and $V_2$ do not have an impact
on the bulk antiferromagnet, as the 
preexisting antiferromagnetic order quenches
other emergent orders, but they only give rise to a finite effect on
the interface states as they are originally gapless. Furthermore,
we have explicitly verified that the emergent topological superconductivity
also appears when a honeycomb lattice with $U$, $V_1$ and $V_2$ is
solved at the mean-field level, with the antiferromagnetic field
dynamically emerging from the selfconsistent solution.

The spin-orbit coupling term emerging from interactions in $\H^{MF}$
is related to the non-local terms $\chi_{ijss'}$ that involve
second-neighbor hoppings and spin-flips.
In particular, the term $\H^{MF}$ can be decomposed in its even and
odd terms with respect to time-reversal symmetry.
From the time-reversal symmetric component, we can extract
the spin-dependent and spin-independent terms. In particular, we have
verified that our self-consistent solution yields
a spin-dependent time-reversal symmetric part
of $\H^{MF}_{KM}$ that takes the form of a spatially-modulated
Kane-Mele spin-orbit coupling\cite{PhysRevLett.95.226801} term of the form

\begin{equation}
        \H^{MF}_{KM} = 
        i\sum_{\langle \langle \alpha \beta \rangle \rangle} 
        \sigma^y_{s,s'} \lambda \left ( \frac{\mathbf r_\alpha + \mathbf r_\beta}{2} \right )
        \nu_{\alpha \beta} c^\dagger_{\alpha,s} c_{\beta,s'}
\end{equation}
so that
\begin{equation}
        \chi^{KM}_{\alpha \beta ss'} =
        i
        \sigma^y_{s,s'} \lambda \left ( \frac{\mathbf r_\alpha + \mathbf r_\beta}{2} \right )
        \nu_{\alpha\beta} 
\end{equation}
where $\chi^{KM}_{\alpha\beta ss'}$ is the time-reversal symmetric, spin-dependent
component of $\chi_{\alpha \beta ss'}$, $\langle \langle 
\rangle \rangle$ denotes
second neighbors, $\nu_{\alpha \beta}=\pm 1$ for clock-wise and
anticlockwise second-neighbor hopping,
and $\lambda (\mathbf r)$ is the spatial
modulation strength of the selfconsistent profile.
We have verified that this term is the one responsible for the
gap opening in our system, whereas the rest of $\H^{MF}$
just creates small renormalizations in the band dispersion.

We now summarize the relation between the gap opening
and the first and second neighbor interactions. The reason
why $V_2$ is capable of opening a gap but not $V_1$ simply stems from
the functional form of those solitonic modes. In particular,
due to the original $U(1)$ spin symmetry of the system, gapping out
the modes require creating spin-mixing between the two solitonic sectors.
However, the interactions parametrized by $V_1$ that could give
rise to spin-mixing yield mean-field single-particle terms that
are zero when evaluated in the solitonic basis. This can be verified
by explicitly adding a Rashba-like spin-orbit coupling
term to the Hamiltonian (the interaction driven spin-mixing term
that could appear from $V_1$), and observing that the
interface modes remain gapless. In stark contrast, the interaction term
parametrized by $V_2$ can potentially lead to a spin-mixing term
of the Kane-Mele form, that when evaluated in the solitonic basis
gives rise to a finite gap opening as explained above.

\section{Estimate of $V_2$ in real materials}
\label{sec:v2}
Here we comment on the expected values of $V_2$ for real materials.
First, it is worth to note that, as shown in our main manuscript,
a strong $V_2 > t$ is not necessary
for the topological phase to appear. 
Actually, we found that a topological gap appears for arbitrarily small
$V_2$. Of course, the bigger the value of $V_2$, the bigger the
topological gap would be. 

In typical two-dimensional materials, second neighbor interactions
are often comparable to the hopping. 
For example, in the case
of graphene, second neighbor interaction is on the order
of $4.2$ eV\cite{PhysRevLett.111.036601}.
This should be compared with
the first neighbor hopping $3$ eV,
giving a ratio 
$V_2/t \approx 1.4$\cite{PhysRevLett.111.036601}. In the case of twisted
two-dimensional materials, known
to host correlated insulating states\cite{Cao2018},
this comparison can become more radical.
For example, twisted graphene bilayers, whose effective model
are also located in a honeycomb lattice\cite{PhysRevX.8.031087},
have second neighbor interactions on the
same order as first neighbor ones\cite{PhysRevX.8.031087}.
A precise quantitative assessment
of the second neighbor interaction in the candidate materials
proposed in our main manuscript
could be performed by exploiting recent developments
in first-principles methods\cite{PhysRevLett.119.056401,PhysRevLett.111.036601,PhysRevB.98.085127,2020arXiv201103271T}.

\bibliography{biblio}{}

\end{document}